\documentclass[10pt]{article}
\usepackage{moriond,epsfig}

\def\be{\begin{equation}}
\def\ee{\end{equation}}
\def\bea{\begin{eqnarray}}
\def\eea{\end{eqnarray}}

\begin{document}
\title{1D Luttinger liquid \& QED}

\author{P.~Degiovanni and S.~Peysson}

\address{Laboratoire de Physique (U.M.R. 5672 du
    CNRS), Ecole Normale Sup\'erieure de Lyon\\46,
  all\'ee d'Italie, 69364 Lyon Cedex 07, France}

\maketitle\abstracts{Using its Conformal Field Theory
  description, the Luttinger liquid is coupled to the
  quantum electromagnetic field. We compute the
  decoherence properties of a superposition of two states
  obtained by creating the same elementary excitation at
  different places in the system.}

\section{Introduction}

The Landau-Fermi theory is the simplest theory for interacting electrons 
in metals.
Its basic low energy excitations called quasiparticles are dressed electrons.
Despite its success in 3D, the 
search for more exotic metallic states has been pursued 
leading to the discovery of the Luttinger liquid. 
This powerful and simple effective theory plays the role of the
Landau-Fermi theory for one dimensional systems and has
many applications.
As illustrated in the present volume, 
various systems such as fractional quantum Hall (FQH) edge states,
carbon nanotubes, quasi-1D organic conductors and quantum
spin chains can be described using this paradigm. 
Up to now, this is the only known effective theory of a non-Fermi liquid.

\medskip

However, the coupling to an external quantum reservoir
such as the quantum electromagnetic field has not attracted much attention.
A notable exception is the work by Loss and Martin \cite{Loss:93-1} 
which shows that QED's
fluctuations
cannot induce a spontaneous permanent
current in a Luttinger ring. A great deal of work has
been done within the context of weak localization but
mainly in the framework of Landau-Fermi theory.
In contrast, questions concerning the real time evolution of a
Luttinger liquid coupled to QED have not been
addressed. 
Here, recent results on 
QED-induced decoherence of Schr\H{o}dinger cat
states in a Luttinger liquid are reported. Our main questions are: 
does QED induce a modification
of the effective Luttinger liquid theory? how strong and
fast is the decoherence induced by the coupling of a
Luttinger liquid to the quantum electromagnetic field?

\medskip

The CFT description of the Luttinger liquid is recalled
in section \ref{sec:LuttCFT}. Our results
concerning decoherence in the Luttinger liquid are
presented in section \ref{secDecoh} 
and the conclusion summarizes the main results.

\section{Effective CFT for the Luttinger Liquid}
\label{sec:LuttCFT}

We consider a system of spinless interacting electrons on a $1D$ 
circle of length $L$.
Such a system can be realized for example using a thin
annular FQH fluid. 

\medskip

As was shown by Haldane \cite{Haldane:81-1} 
for gapless systems\footnote{It means that backscattering terms 
which can potentially open a gap in the spectrum are
discarded. For an annular FQH fluid, the two edges 
sufficiently separated and backscattering is strongly suppressed.}, low
energy properties can be described using an effective
theory with only two parameters: a renormalized Fermi
velocity $v_S$ and a dimensionless interacting constant
$\alpha$. These parameters, which can be related to the
microscopic interaction constants (g-ology) are analogous
of the $f_{\mathbf{k},\mathbf{k}'}$ parameters of
Laudau-Fermi liquids. 

This 1D effective description
can conveniently be formulated in terms of a 2D conformal field
theory \cite{Degio12} (CFT), the Hilbert state of
which can be decomposed according to a
$\widehat{U(1)}_R\times \widehat{U(1)}_L$ symmetry algebra. For each
chirality, $\widehat{U(1)}$ is generated by current modes satisfying: 
$[J_n,J_m] = n\delta_{n,-m} \mathbf{1}$. 
The electric charge and current densities are related to
these modes by:
\bea
\rho(\sigma) &=& \frac{e}{L\sqrt{\alpha}} 
\sum_{l\in \mathbf{Z}} (J_l e^{2i\pi l \sigma /L}
+\bar{J}_l e^{-2i\pi l \sigma /L} )\\
j(\sigma) &=& \frac{ev_S}{L\sqrt{\alpha}} 
\sum_{l\in \mathbf{Z}} (J_l e^{2i\pi l \sigma /L}
-\bar{J}_l e^{-2i\pi l \sigma /L} )
\eea
Irreducible highest weight states of the 
$\widehat{U(1)}_R\times \widehat{U(1)}_L$ symmetry
algebra are associated with conformal primary fields called
vertex operators. The operator content of the effective
Luttinger CFT can be determined by listing these highest
weight states. They are indexed by $(n,m)\in
(\mathbf{Z}/2)\times \mathbf{Z}$ such that $2n\equiv
m\pmod{2}$~: $|n,m\rangle$ and we have:
\begin{equation}
J_0|n,m\rangle =\left(n\sqrt{\alpha}+{m\over
2\sqrt{\alpha}}\right)|n,m\rangle\ \mathrm{and}\quad 
\bar{J_0}|n,m\rangle = \left(n\sqrt{\alpha}-{m\over 2\sqrt{\alpha}}\right)|n,m\rangle
\end{equation}
The $l=0$ modes $J_0$ and $\bar{J}_0$ are quantized:
$2n$ is the total charge and $m$ is related to the total
current circling around the system.
The $l\neq 0$ modes are called {\em  hydrodynamic} 
in reference to Wen's pioneering work \cite{Wen:90-1} on the Luttinger
liquid as a description for edge states
in the FQHE. 

\medskip

A very simple
argument for this identification 
has been proposed \cite{Degio12}. It relies on
the recovery of Laughlin's thought experiment from the CFT point of view
\footnote{Effect of the increase of the magnetic flux
  through the ring on the system's ground state.}. The
Luttinger parameter is related to the filling fraction by
$\alpha\nu =1$. 
From the 1D point of view, the original fermion operators
renormalize (orthogonality catastrophe) to specific
vertex operators: $\psi_R^{\dagger}$
(resp. $\psi_L^{\dagger}$) corresponds to $V_{1/2,1}$ 
(resp. $V_{1/2,-1}$). In the case of a FQH fluid, edge
fermions carrying unit charge on one of the two edges
appear in the spectrum: $V_{1/2,\nu^{-1}}$ and
$V_{1/2,-\nu^{-1}}$. These operators generate an extended
symmetry algebra with respect to which the Luttinger CFT
is rational. In contrast to them, the Luttinger fermions
carry a fractional charge on each edge $q_R = (1+\nu)/2$
and $q_L=(1-\nu)/2$ for $V_{1/2,1}$.

\section{Coupling to quantum electrodynamics}
\label{secDecoh}

In the Coulomb gauge, the Hamiltonian for a
matter system coupled $\mathcal{S}$ to the
electromagnetic 
field $\mathcal{E}$ is given by:
\be
H=H_{\mathcal{S}}(\rho, j) + 
H_{\mathcal{E}}(A_\perp, E_\perp) - \int A_\perp(r) .j_\perp (r)\,d^3r + 
E_\mathrm{Coulomb}
\ee
The static Coulomb interaction is taken into account by the
{\em effective} Luttinger CFT.
In the following, we are interested by the evolution
of the reduced density matrix of the matter system. The
electromagnetic field defines an {\em environment}. 
The Feynman-Vernon-Keldysh \cite{Feynman:1963-1,Keldysh:1965-1} method uses a
path integral language to compute the evolution of $\mathcal{S}$'s
reduced density matrix. Whereas in quantum mechanics the
usual propagator is associated to a simple contour
going from the initial time $t_i$ to the final time
$t_f$, here, the Keldysh contour goes from $t_i$ to
$t_f$ ($+$ branch) and back from $t_f$ to $t_i$
($-$ branch). 
Integration over transverse photons is gaussian 
and immediately leads to the
Feynman-Vernon influence
functional: 
\be
\mathcal{F}[j_+(x),j_-(y)] =
\exp{\left(
-\frac{i}{2}\sum_{{(\epsilon,\epsilon ')\in\{1,-1\}^2\atop 
(k,l)\in \{1,2,3\}^2}}
\epsilon\,\epsilon '\, \int d^4x\,d^4y \,
j^k_{\epsilon}(x)\,
D^{k\,l}_{\epsilon\epsilon '}(x,y)\,
j^l_{\epsilon '}(y)
\right)}
\ee
where the Keldysh-Green functions for the electromagnetic field
$D^{k\,l}_{\epsilon\epsilon '}(x,y)=-i \langle T_K
A ^{k}_{\epsilon}(x)A ^{l}_{\epsilon
  '}(y)\rangle_{\beta}$ should be used\footnote{$\epsilon$
and $\epsilon '$ label the branches of the Keldysh contour.}. 
At this point, the
problem does not seem straightforward~: the effect of the electromagnetic field
is to introduce a non local quartic interaction for the fermions. 

However, such a situation can be handled by the bosonization technique.
We replace fermionic degrees
of freedom by charge and current densities, which are
bosonic and free. In the one dimensional case, this
provides an exact reformulation of the Luttinger
effective theory. Using bosonic coordinates and keeping
only long wavelength components of the current, the coupled
problem Luttinger~\&~QED is gaussian and therefore, in
principle, exactly solvable. 
The method could in principle be extended in $D\geq 2$ but
in this case, bosonization only provides an approximate
description of the fermion system.

\medskip

Using a cylindrical mode decomposition for the
electromagnetic field and the usual mode expansion
for the Luttinger liquid, the problem becomes equivalent to the
linear coupling of an
infinite family of harmonic oscillators with 
distinct baths of oscillators. The underlying elementary
problem (usually called ``QBM'') has been extensively 
studied \cite{CL-83a,HPZ1}.
All dynamical properties are encoded 
into an infinite family of {\em dimensionless spectral densities} 
$\mathcal{J}_l(\omega)$ indexed by Luttinger mode numbers $l\geq 0$. 
At low frequencies, its
behaviour is ohmic for $l=1$ and supraohmic for any other
value: $\mathcal{J}_l(\omega)\simeq (L\omega/c)^{2l-1}$ for
$l\neq 0$ and $\mathcal{J}_0(\omega)\simeq (L\omega/c)^3$. 
The interaction between the Luttinger liquid and the
quantum electromagnetic field is characterized by a 
dimensionless coupling
constant\footnote{$\alpha_{\mathrm{QED}}$ denotes the
  fine structure constant.}:
\be
g=4\pi\,\frac{\alpha_\mathrm{QED}}{\alpha} 
\left( \frac{v_S}{c} \right)^2 \simeq 10^{-8}
\ee
We are interested by the real time evolution of a
state obtained by coherent superposition of two
elementary excitations of the Luttinger liquid.
As recalled in section \ref{sec:LuttCFT}, localized excitations of the
Luttinger liquid are created by vertex operators 
$V_{n,m}(\sigma)$.
Let us consider
Schr\H{o}dinger cat states defined as superpositions of the
same excitation (with possible different chiralities) 
at different places around the circle. For example:
\begin{equation}
\label{defCat}
|\psi_{R/R}\rangle = \frac{1}{\sqrt{2}}\left(
\psi_R^{\dagger}(\sigma_1)|0\rangle 
+ \psi_R^{\dagger}(\sigma_2)|0\rangle 
\right)
\ \mathrm{and}\quad 
|\psi_{R/L}\rangle =\frac{1}{\sqrt{2}}\left(
\psi_R^{\dagger}(\sigma_1)|0\rangle 
+ \psi_L^{\dagger}(\sigma_2)|0\rangle 
\right)
\end{equation}
In an isolated system, such a coherent superposition will remain
coherent. 
Switching on the coupling to the quantum electromagnetic
field changes the situation: according to general works
on decoherence \cite{Zurek:91-1}, such
Schr\H{o}dinger cats should decohere into a
statistical mixing of two states: one excitation at one
position, or the excitation at the other place. From a
solid state physics point of view, decoherence implies that
interference effects are suppressed \cite{Stern:90-1} and
transport leaves the purely quantum regime.

\medskip

Our analysis shows that the
$l=0$ modes can be treated by exact
diagonalization. Hydrodynamic modes are
more complicated to study. But since vertex operators create
coherent states in the $l\geq 1$ modes, we have been
able to obtain the following results.

\paragraph{Zero modes}
Only the total current around the
Luttinger ring couples to the tranverse electric modes.

First, the Luttinger paramaters $v_S$ and $\alpha$ get
renormalized, leading to a modification of the
highest weight state energies. This renormalisation
process takes place within the cutoff time, corresponding
to the UV cutoff frequency $\Lambda $.
Writing $v'_S = v_S \eta$ and $\alpha'=\alpha/\eta$, we have:
\be
\eta^2=1-{c\over \pi v_S}\int_0^\infty 
\mathcal{J}_0(\omega)\,{d\omega \over \omega}
\ee
which is a correction of order $gc/v_S\simeq 10^{-5}$.
Next, off diagonal matrix elements (with respect to $m$)
are damped within the same time scale by a factor:
\begin{equation}
\exp{\left( -\frac{t^2}{(L/c)\alpha}(m-m')^2
\int_0^{+\infty} d\omega \,\mathcal{J}_0(\omega)\,
\frac{1-\cos(\omega t)}{(\omega t)^2}\right)}
\end{equation}
The $t\rightarrow +\infty$ value of the decoherence
exponent is $-{(m-m')^2\over \alpha}\int _0^{+\infty}
{c\over L\omega^2}\,\mathcal{J}_0(\omega)\,d\omega$
which is of typical order $g$. It is proportional to
the square of the difference between current intensities,
a quantity which measures the ``distance'' betweeen the
two quantum states. Since $g\simeq 10^{-8}$, low energy zero modes 
do not loose their quantum
character because of the coupling with QED.

\paragraph{Hydrodynamic modes}
Wigner functions are quite appropriate
to study the evolution of coherent
states. Using the evolution kernel for the evolution of
the density matrix, the exact time
evolution of the Wigner function corresponding to each
hydrodynamic Luttinger mode ($l\geq 1$) can be obtained. In principle,
this method gives access to the short time regime
of the system $t\simeq \Lambda^{-1}$ 
(when non-markovian effects are still
important) and could also be used to study strong
coupling situations. However, due to the weakness of QED's coupling, 
the main features can be found in a simple way.
First of all, let us mention that
energy dissipation into the electromagnetic field
takes place within a time $g^{-1}$ longer than the
typical Luttinger time $L/v_S$. The system is strongly underdamped.

\medskip

In the intermediate regime, $\Lambda^{-1}\ll t \ll
g^{-1}L/v_S$, a perturbative treatment and a
secular approximation enables us to extract terms linear
in time for the decoherence exponent. 
Decoherence times for each of the
Luttinger modes can then be obtained. The main result of this analysis, which
relies on the estimates of low frequency
asympotics of spectral densities
$\mathcal{J}_l(\omega)$ is that the $l=1$ mode dominates
the decoherence process. At zero temperature, the typical decoherence time
constants $\gamma_l$ for the $l$th Luttinger modes are
given by:
\begin{eqnarray}
{L\over v_S}\ldotp  \gamma_l^{(R/R)}(\sigma_1,\sigma_2) & = &  4 g\,
\Delta_{n,m}\left({2\pi v_s\over c}\right)^{2(l-1)}\ldotp 
{l^2(l+1)\over (2l+1)!}\,\ldotp
\sin^2{\left({\pi l\sigma_{12}\over L}\right)}\nonumber
\\
{L\over v_S}\ldotp
  \gamma_l^{(R/L)}(\sigma_1,\sigma_2)
& = & 4g\, \Delta_{n,m}\left({2\pi v_s\over c}\right)^{2(l-1)}\ldotp 
{l^2(l+1)\over (2l+1)!}\,\ldotp 
\left(1+{m^2-4\alpha^2 n^2
\over m^2+4\alpha^2 n^2}
\cos{\left({2\pi l\sigma_{12}\over L}\right)}
\right)\nonumber
\end{eqnarray}
where $\Delta_{n,m}$ denotes the conformal dimension of $V_{n,m}(\sigma)$.
Then $\gamma_{l+1}/\gamma_l \simeq (v_S/c)^2$, 
which shows the dominance of the $l=1$ mode. Results for the
decoherence time of all hydrodynamic modes are presented here for a circle of
radius $5~\mu m$. 
The Luttinger parameter are $\alpha = 3$ and $v_S/c=10^{-3}$. The temperature
dependance can also be obtained: decoherence is faster
as temperature increases. Let us notice that, as could be
expected, the $R/R$ Schr\H{o}dinger cat does not decohere
for $\sigma_{12}=\sigma_1-\sigma_2=0$.

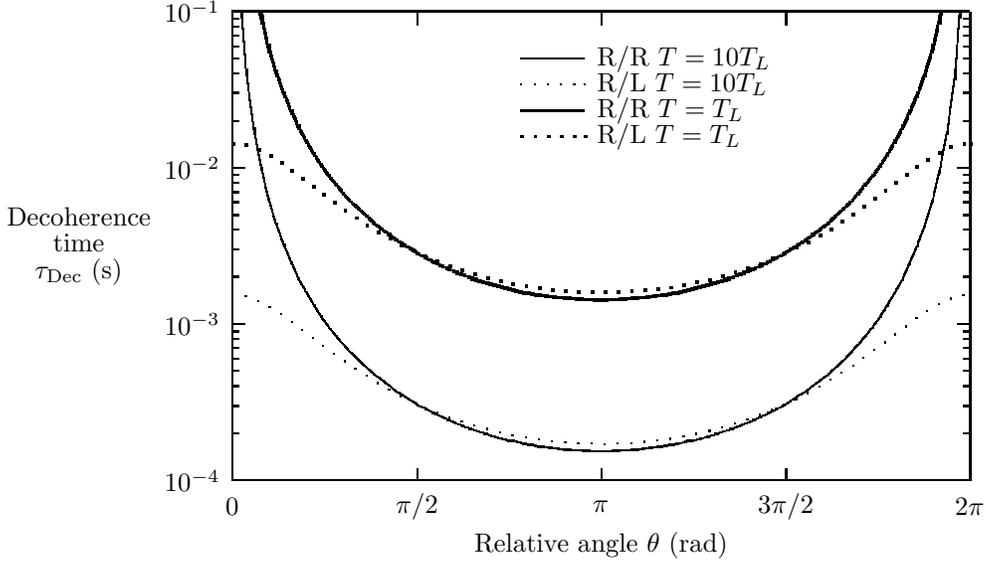
\begin{figure}
\begin{center}
\setlength{\unitlength}{0.240900pt}
\ifx\plotpoint\undefined\newsavebox{\plotpoint}\fi
\sbox{\plotpoint}{\rule[-0.200pt]{0.400pt}{0.400pt}}%
\begin{picture}(1500,900)(0,0)
\font\gnuplot=cmr10 at 10pt
\gnuplot
\sbox{\plotpoint}{\rule[-0.200pt]{0.400pt}{0.400pt}}%
\put(281.0,123.0){\rule[-0.200pt]{4.818pt}{0.400pt}}
\put(261,123){\makebox(0,0)[r]{$10^{-4}$}}
\put(1419.0,123.0){\rule[-0.200pt]{4.818pt}{0.400pt}}
\put(281.0,197.0){\rule[-0.200pt]{2.409pt}{0.400pt}}
\put(1429.0,197.0){\rule[-0.200pt]{2.409pt}{0.400pt}}
\put(281.0,240.0){\rule[-0.200pt]{2.409pt}{0.400pt}}
\put(1429.0,240.0){\rule[-0.200pt]{2.409pt}{0.400pt}}
\put(281.0,271.0){\rule[-0.200pt]{2.409pt}{0.400pt}}
\put(1429.0,271.0){\rule[-0.200pt]{2.409pt}{0.400pt}}
\put(281.0,295.0){\rule[-0.200pt]{2.409pt}{0.400pt}}
\put(1429.0,295.0){\rule[-0.200pt]{2.409pt}{0.400pt}}
\put(281.0,314.0){\rule[-0.200pt]{2.409pt}{0.400pt}}
\put(1429.0,314.0){\rule[-0.200pt]{2.409pt}{0.400pt}}
\put(281.0,331.0){\rule[-0.200pt]{2.409pt}{0.400pt}}
\put(1429.0,331.0){\rule[-0.200pt]{2.409pt}{0.400pt}}
\put(281.0,345.0){\rule[-0.200pt]{2.409pt}{0.400pt}}
\put(1429.0,345.0){\rule[-0.200pt]{2.409pt}{0.400pt}}
\put(281.0,357.0){\rule[-0.200pt]{2.409pt}{0.400pt}}
\put(1429.0,357.0){\rule[-0.200pt]{2.409pt}{0.400pt}}
\put(281.0,369.0){\rule[-0.200pt]{4.818pt}{0.400pt}}
\put(261,369){\makebox(0,0)[r]{$10^{-3}$}}
\put(1419.0,369.0){\rule[-0.200pt]{4.818pt}{0.400pt}}
\put(281.0,443.0){\rule[-0.200pt]{2.409pt}{0.400pt}}
\put(1429.0,443.0){\rule[-0.200pt]{2.409pt}{0.400pt}}
\put(281.0,486.0){\rule[-0.200pt]{2.409pt}{0.400pt}}
\put(1429.0,486.0){\rule[-0.200pt]{2.409pt}{0.400pt}}
\put(281.0,517.0){\rule[-0.200pt]{2.409pt}{0.400pt}}
\put(1429.0,517.0){\rule[-0.200pt]{2.409pt}{0.400pt}}
\put(281.0,540.0){\rule[-0.200pt]{2.409pt}{0.400pt}}
\put(1429.0,540.0){\rule[-0.200pt]{2.409pt}{0.400pt}}
\put(281.0,560.0){\rule[-0.200pt]{2.409pt}{0.400pt}}
\put(1429.0,560.0){\rule[-0.200pt]{2.409pt}{0.400pt}}
\put(281.0,576.0){\rule[-0.200pt]{2.409pt}{0.400pt}}
\put(1429.0,576.0){\rule[-0.200pt]{2.409pt}{0.400pt}}
\put(281.0,591.0){\rule[-0.200pt]{2.409pt}{0.400pt}}
\put(1429.0,591.0){\rule[-0.200pt]{2.409pt}{0.400pt}}
\put(281.0,603.0){\rule[-0.200pt]{2.409pt}{0.400pt}}
\put(1429.0,603.0){\rule[-0.200pt]{2.409pt}{0.400pt}}
\put(281.0,614.0){\rule[-0.200pt]{4.818pt}{0.400pt}}
\put(261,614){\makebox(0,0)[r]{$10^{-2}$}}
\put(1419.0,614.0){\rule[-0.200pt]{4.818pt}{0.400pt}}
\put(281.0,688.0){\rule[-0.200pt]{2.409pt}{0.400pt}}
\put(1429.0,688.0){\rule[-0.200pt]{2.409pt}{0.400pt}}
\put(281.0,732.0){\rule[-0.200pt]{2.409pt}{0.400pt}}
\put(1429.0,732.0){\rule[-0.200pt]{2.409pt}{0.400pt}}
\put(281.0,762.0){\rule[-0.200pt]{2.409pt}{0.400pt}}
\put(1429.0,762.0){\rule[-0.200pt]{2.409pt}{0.400pt}}
\put(281.0,786.0){\rule[-0.200pt]{2.409pt}{0.400pt}}
\put(1429.0,786.0){\rule[-0.200pt]{2.409pt}{0.400pt}}
\put(281.0,805.0){\rule[-0.200pt]{2.409pt}{0.400pt}}
\put(1429.0,805.0){\rule[-0.200pt]{2.409pt}{0.400pt}}
\put(281.0,822.0){\rule[-0.200pt]{2.409pt}{0.400pt}}
\put(1429.0,822.0){\rule[-0.200pt]{2.409pt}{0.400pt}}
\put(281.0,836.0){\rule[-0.200pt]{2.409pt}{0.400pt}}
\put(1429.0,836.0){\rule[-0.200pt]{2.409pt}{0.400pt}}
\put(281.0,849.0){\rule[-0.200pt]{2.409pt}{0.400pt}}
\put(1429.0,849.0){\rule[-0.200pt]{2.409pt}{0.400pt}}
\put(281.0,860.0){\rule[-0.200pt]{4.818pt}{0.400pt}}
\put(261,860){\makebox(0,0)[r]{$10^{-1}$}}
\put(1419.0,860.0){\rule[-0.200pt]{4.818pt}{0.400pt}}
\put(281.0,123.0){\rule[-0.200pt]{0.400pt}{4.818pt}}
\put(281,82){\makebox(0,0){$0$}}
\put(281.0,840.0){\rule[-0.200pt]{0.400pt}{4.818pt}}
\put(571.0,123.0){\rule[-0.200pt]{0.400pt}{4.818pt}}
\put(571,82){\makebox(0,0){$\pi/2$}}
\put(571.0,840.0){\rule[-0.200pt]{0.400pt}{4.818pt}}
\put(860.0,123.0){\rule[-0.200pt]{0.400pt}{4.818pt}}
\put(860,82){\makebox(0,0){$\pi$}}
\put(860.0,840.0){\rule[-0.200pt]{0.400pt}{4.818pt}}
\put(1150.0,123.0){\rule[-0.200pt]{0.400pt}{4.818pt}}
\put(1150,82){\makebox(0,0){$3\pi/2$}}
\put(1150.0,840.0){\rule[-0.200pt]{0.400pt}{4.818pt}}
\put(1439.0,123.0){\rule[-0.200pt]{0.400pt}{4.818pt}}
\put(1439,82){\makebox(0,0){$2\pi$}}
\put(1439.0,840.0){\rule[-0.200pt]{0.400pt}{4.818pt}}
\put(281.0,123.0){\rule[-0.200pt]{278.962pt}{0.400pt}}
\put(1439.0,123.0){\rule[-0.200pt]{0.400pt}{177.543pt}}
\put(281.0,860.0){\rule[-0.200pt]{278.962pt}{0.400pt}}
\put(40,491){\makebox(0,0){\shortstack{Decoherence\\time\\$\tau_\mathrm{Dec}$ (s)}}}
\put(860,21){\makebox(0,0){Relative angle $\theta$ (rad)}}
\put(281.0,123.0){\rule[-0.200pt]{0.400pt}{177.543pt}}
\put(854,786){\makebox(0,0)[l]{R/R $T=10T_L$ }}
\multiput(296.59,838.41)(0.488,-6.701){13}{\rule{0.117pt}{5.200pt}}
\multiput(295.17,849.21)(8.000,-91.207){2}{\rule{0.400pt}{2.600pt}}
\multiput(304.58,745.55)(0.492,-3.727){21}{\rule{0.119pt}{3.000pt}}
\multiput(303.17,751.77)(12.000,-80.773){2}{\rule{0.400pt}{1.500pt}}
\multiput(316.58,662.14)(0.492,-2.607){21}{\rule{0.119pt}{2.133pt}}
\multiput(315.17,666.57)(12.000,-56.572){2}{\rule{0.400pt}{1.067pt}}
\multiput(328.58,602.49)(0.492,-2.194){19}{\rule{0.118pt}{1.809pt}}
\multiput(327.17,606.25)(11.000,-43.245){2}{\rule{0.400pt}{0.905pt}}
\multiput(339.58,557.19)(0.492,-1.659){21}{\rule{0.119pt}{1.400pt}}
\multiput(338.17,560.09)(12.000,-36.094){2}{\rule{0.400pt}{0.700pt}}
\multiput(351.58,519.16)(0.492,-1.358){21}{\rule{0.119pt}{1.167pt}}
\multiput(350.17,521.58)(12.000,-29.579){2}{\rule{0.400pt}{0.583pt}}
\multiput(363.58,487.71)(0.492,-1.186){21}{\rule{0.119pt}{1.033pt}}
\multiput(362.17,489.86)(12.000,-25.855){2}{\rule{0.400pt}{0.517pt}}
\multiput(375.58,459.81)(0.492,-1.156){19}{\rule{0.118pt}{1.009pt}}
\multiput(374.17,461.91)(11.000,-22.906){2}{\rule{0.400pt}{0.505pt}}
\multiput(386.58,435.68)(0.492,-0.884){21}{\rule{0.119pt}{0.800pt}}
\multiput(385.17,437.34)(12.000,-19.340){2}{\rule{0.400pt}{0.400pt}}
\multiput(398.58,414.82)(0.492,-0.841){21}{\rule{0.119pt}{0.767pt}}
\multiput(397.17,416.41)(12.000,-18.409){2}{\rule{0.400pt}{0.383pt}}
\multiput(410.58,394.87)(0.492,-0.826){19}{\rule{0.118pt}{0.755pt}}
\multiput(409.17,396.43)(11.000,-16.434){2}{\rule{0.400pt}{0.377pt}}
\multiput(421.58,377.37)(0.492,-0.669){21}{\rule{0.119pt}{0.633pt}}
\multiput(420.17,378.69)(12.000,-14.685){2}{\rule{0.400pt}{0.317pt}}
\multiput(433.58,361.51)(0.492,-0.625){21}{\rule{0.119pt}{0.600pt}}
\multiput(432.17,362.75)(12.000,-13.755){2}{\rule{0.400pt}{0.300pt}}
\multiput(445.58,346.62)(0.492,-0.590){19}{\rule{0.118pt}{0.573pt}}
\multiput(444.17,347.81)(11.000,-11.811){2}{\rule{0.400pt}{0.286pt}}
\multiput(456.58,333.79)(0.492,-0.539){21}{\rule{0.119pt}{0.533pt}}
\multiput(455.17,334.89)(12.000,-11.893){2}{\rule{0.400pt}{0.267pt}}
\multiput(468.00,321.92)(0.496,-0.492){21}{\rule{0.500pt}{0.119pt}}
\multiput(468.00,322.17)(10.962,-12.000){2}{\rule{0.250pt}{0.400pt}}
\multiput(480.00,309.92)(0.543,-0.492){19}{\rule{0.536pt}{0.118pt}}
\multiput(480.00,310.17)(10.887,-11.000){2}{\rule{0.268pt}{0.400pt}}
\multiput(492.00,298.92)(0.547,-0.491){17}{\rule{0.540pt}{0.118pt}}
\multiput(492.00,299.17)(9.879,-10.000){2}{\rule{0.270pt}{0.400pt}}
\multiput(503.00,288.93)(0.669,-0.489){15}{\rule{0.633pt}{0.118pt}}
\multiput(503.00,289.17)(10.685,-9.000){2}{\rule{0.317pt}{0.400pt}}
\multiput(515.00,279.93)(0.669,-0.489){15}{\rule{0.633pt}{0.118pt}}
\multiput(515.00,280.17)(10.685,-9.000){2}{\rule{0.317pt}{0.400pt}}
\multiput(527.00,270.93)(0.611,-0.489){15}{\rule{0.589pt}{0.118pt}}
\multiput(527.00,271.17)(9.778,-9.000){2}{\rule{0.294pt}{0.400pt}}
\multiput(538.00,261.93)(0.874,-0.485){11}{\rule{0.786pt}{0.117pt}}
\multiput(538.00,262.17)(10.369,-7.000){2}{\rule{0.393pt}{0.400pt}}
\multiput(550.00,254.93)(0.758,-0.488){13}{\rule{0.700pt}{0.117pt}}
\multiput(550.00,255.17)(10.547,-8.000){2}{\rule{0.350pt}{0.400pt}}
\multiput(562.00,246.93)(0.798,-0.485){11}{\rule{0.729pt}{0.117pt}}
\multiput(562.00,247.17)(9.488,-7.000){2}{\rule{0.364pt}{0.400pt}}
\multiput(573.00,239.93)(1.033,-0.482){9}{\rule{0.900pt}{0.116pt}}
\multiput(573.00,240.17)(10.132,-6.000){2}{\rule{0.450pt}{0.400pt}}
\multiput(585.00,233.93)(1.033,-0.482){9}{\rule{0.900pt}{0.116pt}}
\multiput(585.00,234.17)(10.132,-6.000){2}{\rule{0.450pt}{0.400pt}}
\multiput(597.00,227.93)(1.033,-0.482){9}{\rule{0.900pt}{0.116pt}}
\multiput(597.00,228.17)(10.132,-6.000){2}{\rule{0.450pt}{0.400pt}}
\multiput(609.00,221.93)(1.155,-0.477){7}{\rule{0.980pt}{0.115pt}}
\multiput(609.00,222.17)(8.966,-5.000){2}{\rule{0.490pt}{0.400pt}}
\multiput(620.00,216.93)(1.267,-0.477){7}{\rule{1.060pt}{0.115pt}}
\multiput(620.00,217.17)(9.800,-5.000){2}{\rule{0.530pt}{0.400pt}}
\multiput(632.00,211.93)(1.267,-0.477){7}{\rule{1.060pt}{0.115pt}}
\multiput(632.00,212.17)(9.800,-5.000){2}{\rule{0.530pt}{0.400pt}}
\multiput(644.00,206.94)(1.505,-0.468){5}{\rule{1.200pt}{0.113pt}}
\multiput(644.00,207.17)(8.509,-4.000){2}{\rule{0.600pt}{0.400pt}}
\multiput(655.00,202.94)(1.651,-0.468){5}{\rule{1.300pt}{0.113pt}}
\multiput(655.00,203.17)(9.302,-4.000){2}{\rule{0.650pt}{0.400pt}}
\multiput(667.00,198.94)(1.651,-0.468){5}{\rule{1.300pt}{0.113pt}}
\multiput(667.00,199.17)(9.302,-4.000){2}{\rule{0.650pt}{0.400pt}}
\multiput(679.00,194.94)(1.505,-0.468){5}{\rule{1.200pt}{0.113pt}}
\multiput(679.00,195.17)(8.509,-4.000){2}{\rule{0.600pt}{0.400pt}}
\multiput(690.00,190.95)(2.472,-0.447){3}{\rule{1.700pt}{0.108pt}}
\multiput(690.00,191.17)(8.472,-3.000){2}{\rule{0.850pt}{0.400pt}}
\multiput(702.00,187.95)(2.472,-0.447){3}{\rule{1.700pt}{0.108pt}}
\multiput(702.00,188.17)(8.472,-3.000){2}{\rule{0.850pt}{0.400pt}}
\put(714,184.17){\rule{2.300pt}{0.400pt}}
\multiput(714.00,185.17)(6.226,-2.000){2}{\rule{1.150pt}{0.400pt}}
\multiput(725.00,182.95)(2.472,-0.447){3}{\rule{1.700pt}{0.108pt}}
\multiput(725.00,183.17)(8.472,-3.000){2}{\rule{0.850pt}{0.400pt}}
\put(737,179.17){\rule{2.500pt}{0.400pt}}
\multiput(737.00,180.17)(6.811,-2.000){2}{\rule{1.250pt}{0.400pt}}
\put(749,177.17){\rule{2.500pt}{0.400pt}}
\multiput(749.00,178.17)(6.811,-2.000){2}{\rule{1.250pt}{0.400pt}}
\put(761,175.17){\rule{2.300pt}{0.400pt}}
\multiput(761.00,176.17)(6.226,-2.000){2}{\rule{1.150pt}{0.400pt}}
\put(772,173.67){\rule{2.891pt}{0.400pt}}
\multiput(772.00,174.17)(6.000,-1.000){2}{\rule{1.445pt}{0.400pt}}
\put(784,172.17){\rule{2.500pt}{0.400pt}}
\multiput(784.00,173.17)(6.811,-2.000){2}{\rule{1.250pt}{0.400pt}}
\put(796,170.67){\rule{2.650pt}{0.400pt}}
\multiput(796.00,171.17)(5.500,-1.000){2}{\rule{1.325pt}{0.400pt}}
\put(807,169.67){\rule{2.891pt}{0.400pt}}
\multiput(807.00,170.17)(6.000,-1.000){2}{\rule{1.445pt}{0.400pt}}
\put(734.0,786.0){\rule[-0.200pt]{24.090pt}{0.400pt}}
\put(831,168.67){\rule{2.650pt}{0.400pt}}
\multiput(831.00,169.17)(5.500,-1.000){2}{\rule{1.325pt}{0.400pt}}
\put(819.0,170.0){\rule[-0.200pt]{2.891pt}{0.400pt}}
\put(878,168.67){\rule{2.650pt}{0.400pt}}
\multiput(878.00,168.17)(5.500,1.000){2}{\rule{1.325pt}{0.400pt}}
\put(842.0,169.0){\rule[-0.200pt]{8.672pt}{0.400pt}}
\put(901,169.67){\rule{2.891pt}{0.400pt}}
\multiput(901.00,169.17)(6.000,1.000){2}{\rule{1.445pt}{0.400pt}}
\put(913,170.67){\rule{2.650pt}{0.400pt}}
\multiput(913.00,170.17)(5.500,1.000){2}{\rule{1.325pt}{0.400pt}}
\put(924,172.17){\rule{2.500pt}{0.400pt}}
\multiput(924.00,171.17)(6.811,2.000){2}{\rule{1.250pt}{0.400pt}}
\put(936,173.67){\rule{2.891pt}{0.400pt}}
\multiput(936.00,173.17)(6.000,1.000){2}{\rule{1.445pt}{0.400pt}}
\put(948,175.17){\rule{2.300pt}{0.400pt}}
\multiput(948.00,174.17)(6.226,2.000){2}{\rule{1.150pt}{0.400pt}}
\put(959,177.17){\rule{2.500pt}{0.400pt}}
\multiput(959.00,176.17)(6.811,2.000){2}{\rule{1.250pt}{0.400pt}}
\put(971,179.17){\rule{2.500pt}{0.400pt}}
\multiput(971.00,178.17)(6.811,2.000){2}{\rule{1.250pt}{0.400pt}}
\multiput(983.00,181.61)(2.472,0.447){3}{\rule{1.700pt}{0.108pt}}
\multiput(983.00,180.17)(8.472,3.000){2}{\rule{0.850pt}{0.400pt}}
\put(995,184.17){\rule{2.300pt}{0.400pt}}
\multiput(995.00,183.17)(6.226,2.000){2}{\rule{1.150pt}{0.400pt}}
\multiput(1006.00,186.61)(2.472,0.447){3}{\rule{1.700pt}{0.108pt}}
\multiput(1006.00,185.17)(8.472,3.000){2}{\rule{0.850pt}{0.400pt}}
\multiput(1018.00,189.61)(2.472,0.447){3}{\rule{1.700pt}{0.108pt}}
\multiput(1018.00,188.17)(8.472,3.000){2}{\rule{0.850pt}{0.400pt}}
\multiput(1030.00,192.60)(1.505,0.468){5}{\rule{1.200pt}{0.113pt}}
\multiput(1030.00,191.17)(8.509,4.000){2}{\rule{0.600pt}{0.400pt}}
\multiput(1041.00,196.60)(1.651,0.468){5}{\rule{1.300pt}{0.113pt}}
\multiput(1041.00,195.17)(9.302,4.000){2}{\rule{0.650pt}{0.400pt}}
\multiput(1053.00,200.60)(1.651,0.468){5}{\rule{1.300pt}{0.113pt}}
\multiput(1053.00,199.17)(9.302,4.000){2}{\rule{0.650pt}{0.400pt}}
\multiput(1065.00,204.60)(1.505,0.468){5}{\rule{1.200pt}{0.113pt}}
\multiput(1065.00,203.17)(8.509,4.000){2}{\rule{0.600pt}{0.400pt}}
\multiput(1076.00,208.59)(1.267,0.477){7}{\rule{1.060pt}{0.115pt}}
\multiput(1076.00,207.17)(9.800,5.000){2}{\rule{0.530pt}{0.400pt}}
\multiput(1088.00,213.59)(1.267,0.477){7}{\rule{1.060pt}{0.115pt}}
\multiput(1088.00,212.17)(9.800,5.000){2}{\rule{0.530pt}{0.400pt}}
\multiput(1100.00,218.59)(1.155,0.477){7}{\rule{0.980pt}{0.115pt}}
\multiput(1100.00,217.17)(8.966,5.000){2}{\rule{0.490pt}{0.400pt}}
\multiput(1111.00,223.59)(1.033,0.482){9}{\rule{0.900pt}{0.116pt}}
\multiput(1111.00,222.17)(10.132,6.000){2}{\rule{0.450pt}{0.400pt}}
\multiput(1123.00,229.59)(1.033,0.482){9}{\rule{0.900pt}{0.116pt}}
\multiput(1123.00,228.17)(10.132,6.000){2}{\rule{0.450pt}{0.400pt}}
\multiput(1135.00,235.59)(1.033,0.482){9}{\rule{0.900pt}{0.116pt}}
\multiput(1135.00,234.17)(10.132,6.000){2}{\rule{0.450pt}{0.400pt}}
\multiput(1147.00,241.59)(0.798,0.485){11}{\rule{0.729pt}{0.117pt}}
\multiput(1147.00,240.17)(9.488,7.000){2}{\rule{0.364pt}{0.400pt}}
\multiput(1158.00,248.59)(0.758,0.488){13}{\rule{0.700pt}{0.117pt}}
\multiput(1158.00,247.17)(10.547,8.000){2}{\rule{0.350pt}{0.400pt}}
\multiput(1170.00,256.59)(0.874,0.485){11}{\rule{0.786pt}{0.117pt}}
\multiput(1170.00,255.17)(10.369,7.000){2}{\rule{0.393pt}{0.400pt}}
\multiput(1182.00,263.59)(0.611,0.489){15}{\rule{0.589pt}{0.118pt}}
\multiput(1182.00,262.17)(9.778,9.000){2}{\rule{0.294pt}{0.400pt}}
\multiput(1193.00,272.59)(0.669,0.489){15}{\rule{0.633pt}{0.118pt}}
\multiput(1193.00,271.17)(10.685,9.000){2}{\rule{0.317pt}{0.400pt}}
\multiput(1205.00,281.59)(0.669,0.489){15}{\rule{0.633pt}{0.118pt}}
\multiput(1205.00,280.17)(10.685,9.000){2}{\rule{0.317pt}{0.400pt}}
\multiput(1217.00,290.58)(0.547,0.491){17}{\rule{0.540pt}{0.118pt}}
\multiput(1217.00,289.17)(9.879,10.000){2}{\rule{0.270pt}{0.400pt}}
\multiput(1228.00,300.58)(0.543,0.492){19}{\rule{0.536pt}{0.118pt}}
\multiput(1228.00,299.17)(10.887,11.000){2}{\rule{0.268pt}{0.400pt}}
\multiput(1240.00,311.58)(0.496,0.492){21}{\rule{0.500pt}{0.119pt}}
\multiput(1240.00,310.17)(10.962,12.000){2}{\rule{0.250pt}{0.400pt}}
\multiput(1252.58,323.00)(0.492,0.539){21}{\rule{0.119pt}{0.533pt}}
\multiput(1251.17,323.00)(12.000,11.893){2}{\rule{0.400pt}{0.267pt}}
\multiput(1264.58,336.00)(0.492,0.590){19}{\rule{0.118pt}{0.573pt}}
\multiput(1263.17,336.00)(11.000,11.811){2}{\rule{0.400pt}{0.286pt}}
\multiput(1275.58,349.00)(0.492,0.625){21}{\rule{0.119pt}{0.600pt}}
\multiput(1274.17,349.00)(12.000,13.755){2}{\rule{0.400pt}{0.300pt}}
\multiput(1287.58,364.00)(0.492,0.669){21}{\rule{0.119pt}{0.633pt}}
\multiput(1286.17,364.00)(12.000,14.685){2}{\rule{0.400pt}{0.317pt}}
\multiput(1299.58,380.00)(0.492,0.826){19}{\rule{0.118pt}{0.755pt}}
\multiput(1298.17,380.00)(11.000,16.434){2}{\rule{0.400pt}{0.377pt}}
\multiput(1310.58,398.00)(0.492,0.841){21}{\rule{0.119pt}{0.767pt}}
\multiput(1309.17,398.00)(12.000,18.409){2}{\rule{0.400pt}{0.383pt}}
\multiput(1322.58,418.00)(0.492,0.884){21}{\rule{0.119pt}{0.800pt}}
\multiput(1321.17,418.00)(12.000,19.340){2}{\rule{0.400pt}{0.400pt}}
\multiput(1334.58,439.00)(0.492,1.156){19}{\rule{0.118pt}{1.009pt}}
\multiput(1333.17,439.00)(11.000,22.906){2}{\rule{0.400pt}{0.505pt}}
\multiput(1345.58,464.00)(0.492,1.186){21}{\rule{0.119pt}{1.033pt}}
\multiput(1344.17,464.00)(12.000,25.855){2}{\rule{0.400pt}{0.517pt}}
\multiput(1357.58,492.00)(0.492,1.358){21}{\rule{0.119pt}{1.167pt}}
\multiput(1356.17,492.00)(12.000,29.579){2}{\rule{0.400pt}{0.583pt}}
\multiput(1369.58,524.00)(0.492,1.659){21}{\rule{0.119pt}{1.400pt}}
\multiput(1368.17,524.00)(12.000,36.094){2}{\rule{0.400pt}{0.700pt}}
\multiput(1381.58,563.00)(0.492,2.194){19}{\rule{0.118pt}{1.809pt}}
\multiput(1380.17,563.00)(11.000,43.245){2}{\rule{0.400pt}{0.905pt}}
\multiput(1392.58,610.00)(0.492,2.607){21}{\rule{0.119pt}{2.133pt}}
\multiput(1391.17,610.00)(12.000,56.572){2}{\rule{0.400pt}{1.067pt}}
\multiput(1404.58,671.00)(0.492,3.727){21}{\rule{0.119pt}{3.000pt}}
\multiput(1403.17,671.00)(12.000,80.773){2}{\rule{0.400pt}{1.500pt}}
\multiput(1416.59,758.00)(0.488,6.701){13}{\rule{0.117pt}{5.200pt}}
\multiput(1415.17,758.00)(8.000,91.207){2}{\rule{0.400pt}{2.600pt}}
\put(889.0,170.0){\rule[-0.200pt]{2.891pt}{0.400pt}}
\put(854,745){\makebox(0,0)[l]{R/L $T=10T_L$ }}
\multiput(734,745)(20.756,0.000){5}{\usebox{\plotpoint}}
\put(834,745){\usebox{\plotpoint}}
\put(281,415){\usebox{\plotpoint}}
\put(281.00,415.00){\usebox{\plotpoint}}
\put(301.41,411.71){\usebox{\plotpoint}}
\put(321.00,404.92){\usebox{\plotpoint}}
\put(339.12,394.93){\usebox{\plotpoint}}
\put(356.82,384.12){\usebox{\plotpoint}}
\put(373.67,372.00){\usebox{\plotpoint}}
\put(389.90,359.08){\usebox{\plotpoint}}
\put(406.50,346.62){\usebox{\plotpoint}}
\put(422.81,333.79){\usebox{\plotpoint}}
\put(439.81,321.90){\usebox{\plotpoint}}
\put(456.53,309.60){\usebox{\plotpoint}}
\put(473.54,297.77){\usebox{\plotpoint}}
\put(491.05,286.63){\usebox{\plotpoint}}
\put(508.68,275.69){\usebox{\plotpoint}}
\put(526.61,265.23){\usebox{\plotpoint}}
\put(544.95,255.53){\usebox{\plotpoint}}
\put(563.54,246.30){\usebox{\plotpoint}}
\put(582.27,237.36){\usebox{\plotpoint}}
\put(601.67,230.06){\usebox{\plotpoint}}
\put(620.69,221.77){\usebox{\plotpoint}}
\put(640.38,215.21){\usebox{\plotpoint}}
\put(660.25,209.25){\usebox{\plotpoint}}
\put(680.23,203.66){\usebox{\plotpoint}}
\put(700.31,198.42){\usebox{\plotpoint}}
\put(720.54,193.81){\usebox{\plotpoint}}
\put(741.00,190.33){\usebox{\plotpoint}}
\put(761.48,186.96){\usebox{\plotpoint}}
\put(782.05,184.32){\usebox{\plotpoint}}
\put(802.71,182.39){\usebox{\plotpoint}}
\put(823.41,181.00){\usebox{\plotpoint}}
\put(844.16,180.82){\usebox{\plotpoint}}
\put(864.88,180.00){\usebox{\plotpoint}}
\put(885.59,181.00){\usebox{\plotpoint}}
\put(906.33,181.44){\usebox{\plotpoint}}
\put(927.00,183.25){\usebox{\plotpoint}}
\put(947.57,185.93){\usebox{\plotpoint}}
\put(968.15,188.52){\usebox{\plotpoint}}
\put(988.62,191.94){\usebox{\plotpoint}}
\put(1009.01,195.75){\usebox{\plotpoint}}
\put(1029.15,200.79){\usebox{\plotpoint}}
\put(1049.22,206.06){\usebox{\plotpoint}}
\put(1069.07,212.11){\usebox{\plotpoint}}
\put(1088.87,218.29){\usebox{\plotpoint}}
\put(1108.22,225.74){\usebox{\plotpoint}}
\put(1127.46,233.49){\usebox{\plotpoint}}
\put(1146.45,241.73){\usebox{\plotpoint}}
\put(1165.21,250.60){\usebox{\plotpoint}}
\put(1183.74,259.95){\usebox{\plotpoint}}
\put(1201.82,270.14){\usebox{\plotpoint}}
\put(1219.68,280.71){\usebox{\plotpoint}}
\put(1237.07,292.04){\usebox{\plotpoint}}
\put(1254.67,303.00){\usebox{\plotpoint}}
\put(1271.35,315.35){\usebox{\plotpoint}}
\put(1288.04,327.69){\usebox{\plotpoint}}
\put(1304.87,339.80){\usebox{\plotpoint}}
\put(1321.30,352.47){\usebox{\plotpoint}}
\put(1337.78,365.09){\usebox{\plotpoint}}
\put(1354.14,377.85){\usebox{\plotpoint}}
\put(1371.38,389.39){\usebox{\plotpoint}}
\put(1389.11,400.16){\usebox{\plotpoint}}
\put(1408.11,408.37){\usebox{\plotpoint}}
\put(1428.04,414.09){\usebox{\plotpoint}}
\put(1439,415){\usebox{\plotpoint}}
\sbox{\plotpoint}{\rule[-0.400pt]{0.800pt}{0.800pt}}%
\put(854,704){\makebox(0,0)[l]{R/R $T=T_L$ }}
\put(324.84,847){\rule{0.800pt}{3.132pt}}
\multiput(323.34,853.50)(3.000,-6.500){2}{\rule{0.800pt}{1.566pt}}
\multiput(329.40,831.98)(0.512,-2.284){15}{\rule{0.123pt}{3.618pt}}
\multiput(326.34,839.49)(11.000,-39.490){2}{\rule{0.800pt}{1.809pt}}
\multiput(340.41,788.65)(0.511,-1.666){17}{\rule{0.123pt}{2.733pt}}
\multiput(337.34,794.33)(12.000,-32.327){2}{\rule{0.800pt}{1.367pt}}
\multiput(352.41,752.04)(0.511,-1.440){17}{\rule{0.123pt}{2.400pt}}
\multiput(349.34,757.02)(12.000,-28.019){2}{\rule{0.800pt}{1.200pt}}
\multiput(364.41,720.42)(0.511,-1.214){17}{\rule{0.123pt}{2.067pt}}
\multiput(361.34,724.71)(12.000,-23.711){2}{\rule{0.800pt}{1.033pt}}
\multiput(376.40,692.92)(0.512,-1.137){15}{\rule{0.123pt}{1.945pt}}
\multiput(373.34,696.96)(11.000,-19.962){2}{\rule{0.800pt}{0.973pt}}
\multiput(387.41,670.08)(0.511,-0.943){17}{\rule{0.123pt}{1.667pt}}
\multiput(384.34,673.54)(12.000,-18.541){2}{\rule{0.800pt}{0.833pt}}
\multiput(399.41,648.64)(0.511,-0.852){17}{\rule{0.123pt}{1.533pt}}
\multiput(396.34,651.82)(12.000,-16.817){2}{\rule{0.800pt}{0.767pt}}
\multiput(411.40,629.04)(0.512,-0.788){15}{\rule{0.123pt}{1.436pt}}
\multiput(408.34,632.02)(11.000,-14.019){2}{\rule{0.800pt}{0.718pt}}
\multiput(422.41,612.47)(0.511,-0.717){17}{\rule{0.123pt}{1.333pt}}
\multiput(419.34,615.23)(12.000,-14.233){2}{\rule{0.800pt}{0.667pt}}
\multiput(434.41,596.30)(0.511,-0.581){17}{\rule{0.123pt}{1.133pt}}
\multiput(431.34,598.65)(12.000,-11.648){2}{\rule{0.800pt}{0.567pt}}
\multiput(446.40,581.94)(0.512,-0.639){15}{\rule{0.123pt}{1.218pt}}
\multiput(443.34,584.47)(11.000,-11.472){2}{\rule{0.800pt}{0.609pt}}
\multiput(457.41,568.57)(0.511,-0.536){17}{\rule{0.123pt}{1.067pt}}
\multiput(454.34,570.79)(12.000,-10.786){2}{\rule{0.800pt}{0.533pt}}
\multiput(468.00,558.08)(0.539,-0.512){15}{\rule{1.073pt}{0.123pt}}
\multiput(468.00,558.34)(9.774,-11.000){2}{\rule{0.536pt}{0.800pt}}
\multiput(480.00,547.08)(0.539,-0.512){15}{\rule{1.073pt}{0.123pt}}
\multiput(480.00,547.34)(9.774,-11.000){2}{\rule{0.536pt}{0.800pt}}
\multiput(492.00,536.08)(0.489,-0.512){15}{\rule{1.000pt}{0.123pt}}
\multiput(492.00,536.34)(8.924,-11.000){2}{\rule{0.500pt}{0.800pt}}
\multiput(503.00,525.08)(0.674,-0.516){11}{\rule{1.267pt}{0.124pt}}
\multiput(503.00,525.34)(9.371,-9.000){2}{\rule{0.633pt}{0.800pt}}
\multiput(515.00,516.08)(0.674,-0.516){11}{\rule{1.267pt}{0.124pt}}
\multiput(515.00,516.34)(9.371,-9.000){2}{\rule{0.633pt}{0.800pt}}
\multiput(527.00,507.08)(0.700,-0.520){9}{\rule{1.300pt}{0.125pt}}
\multiput(527.00,507.34)(8.302,-8.000){2}{\rule{0.650pt}{0.800pt}}
\multiput(538.00,499.08)(0.774,-0.520){9}{\rule{1.400pt}{0.125pt}}
\multiput(538.00,499.34)(9.094,-8.000){2}{\rule{0.700pt}{0.800pt}}
\multiput(550.00,491.08)(0.913,-0.526){7}{\rule{1.571pt}{0.127pt}}
\multiput(550.00,491.34)(8.738,-7.000){2}{\rule{0.786pt}{0.800pt}}
\multiput(562.00,484.08)(0.825,-0.526){7}{\rule{1.457pt}{0.127pt}}
\multiput(562.00,484.34)(7.976,-7.000){2}{\rule{0.729pt}{0.800pt}}
\multiput(573.00,477.08)(0.913,-0.526){7}{\rule{1.571pt}{0.127pt}}
\multiput(573.00,477.34)(8.738,-7.000){2}{\rule{0.786pt}{0.800pt}}
\multiput(585.00,470.07)(1.132,-0.536){5}{\rule{1.800pt}{0.129pt}}
\multiput(585.00,470.34)(8.264,-6.000){2}{\rule{0.900pt}{0.800pt}}
\multiput(597.00,464.07)(1.132,-0.536){5}{\rule{1.800pt}{0.129pt}}
\multiput(597.00,464.34)(8.264,-6.000){2}{\rule{0.900pt}{0.800pt}}
\multiput(609.00,458.06)(1.432,-0.560){3}{\rule{1.960pt}{0.135pt}}
\multiput(609.00,458.34)(6.932,-5.000){2}{\rule{0.980pt}{0.800pt}}
\multiput(620.00,453.06)(1.600,-0.560){3}{\rule{2.120pt}{0.135pt}}
\multiput(620.00,453.34)(7.600,-5.000){2}{\rule{1.060pt}{0.800pt}}
\put(632,446.34){\rule{2.600pt}{0.800pt}}
\multiput(632.00,448.34)(6.604,-4.000){2}{\rule{1.300pt}{0.800pt}}
\multiput(644.00,444.06)(1.432,-0.560){3}{\rule{1.960pt}{0.135pt}}
\multiput(644.00,444.34)(6.932,-5.000){2}{\rule{0.980pt}{0.800pt}}
\put(655,437.34){\rule{2.600pt}{0.800pt}}
\multiput(655.00,439.34)(6.604,-4.000){2}{\rule{1.300pt}{0.800pt}}
\put(667,433.34){\rule{2.600pt}{0.800pt}}
\multiput(667.00,435.34)(6.604,-4.000){2}{\rule{1.300pt}{0.800pt}}
\put(679,429.84){\rule{2.650pt}{0.800pt}}
\multiput(679.00,431.34)(5.500,-3.000){2}{\rule{1.325pt}{0.800pt}}
\put(690,426.84){\rule{2.891pt}{0.800pt}}
\multiput(690.00,428.34)(6.000,-3.000){2}{\rule{1.445pt}{0.800pt}}
\put(702,423.84){\rule{2.891pt}{0.800pt}}
\multiput(702.00,425.34)(6.000,-3.000){2}{\rule{1.445pt}{0.800pt}}
\put(714,420.84){\rule{2.650pt}{0.800pt}}
\multiput(714.00,422.34)(5.500,-3.000){2}{\rule{1.325pt}{0.800pt}}
\put(725,417.84){\rule{2.891pt}{0.800pt}}
\multiput(725.00,419.34)(6.000,-3.000){2}{\rule{1.445pt}{0.800pt}}
\put(737,415.34){\rule{2.891pt}{0.800pt}}
\multiput(737.00,416.34)(6.000,-2.000){2}{\rule{1.445pt}{0.800pt}}
\put(749,413.34){\rule{2.891pt}{0.800pt}}
\multiput(749.00,414.34)(6.000,-2.000){2}{\rule{1.445pt}{0.800pt}}
\put(761,411.84){\rule{2.650pt}{0.800pt}}
\multiput(761.00,412.34)(5.500,-1.000){2}{\rule{1.325pt}{0.800pt}}
\put(772,410.34){\rule{2.891pt}{0.800pt}}
\multiput(772.00,411.34)(6.000,-2.000){2}{\rule{1.445pt}{0.800pt}}
\put(784,408.84){\rule{2.891pt}{0.800pt}}
\multiput(784.00,409.34)(6.000,-1.000){2}{\rule{1.445pt}{0.800pt}}
\put(796,407.84){\rule{2.650pt}{0.800pt}}
\multiput(796.00,408.34)(5.500,-1.000){2}{\rule{1.325pt}{0.800pt}}
\put(807,406.84){\rule{2.891pt}{0.800pt}}
\multiput(807.00,407.34)(6.000,-1.000){2}{\rule{1.445pt}{0.800pt}}
\put(819,405.84){\rule{2.891pt}{0.800pt}}
\multiput(819.00,406.34)(6.000,-1.000){2}{\rule{1.445pt}{0.800pt}}
\put(734.0,704.0){\rule[-0.400pt]{24.090pt}{0.800pt}}
\put(842,404.84){\rule{2.891pt}{0.800pt}}
\multiput(842.00,405.34)(6.000,-1.000){2}{\rule{1.445pt}{0.800pt}}
\put(831.0,407.0){\rule[-0.400pt]{2.650pt}{0.800pt}}
\put(866,404.84){\rule{2.891pt}{0.800pt}}
\multiput(866.00,404.34)(6.000,1.000){2}{\rule{1.445pt}{0.800pt}}
\put(854.0,406.0){\rule[-0.400pt]{2.891pt}{0.800pt}}
\put(889,405.84){\rule{2.891pt}{0.800pt}}
\multiput(889.00,405.34)(6.000,1.000){2}{\rule{1.445pt}{0.800pt}}
\put(901,406.84){\rule{2.891pt}{0.800pt}}
\multiput(901.00,406.34)(6.000,1.000){2}{\rule{1.445pt}{0.800pt}}
\put(913,407.84){\rule{2.650pt}{0.800pt}}
\multiput(913.00,407.34)(5.500,1.000){2}{\rule{1.325pt}{0.800pt}}
\put(924,408.84){\rule{2.891pt}{0.800pt}}
\multiput(924.00,408.34)(6.000,1.000){2}{\rule{1.445pt}{0.800pt}}
\put(936,410.34){\rule{2.891pt}{0.800pt}}
\multiput(936.00,409.34)(6.000,2.000){2}{\rule{1.445pt}{0.800pt}}
\put(948,411.84){\rule{2.650pt}{0.800pt}}
\multiput(948.00,411.34)(5.500,1.000){2}{\rule{1.325pt}{0.800pt}}
\put(959,413.34){\rule{2.891pt}{0.800pt}}
\multiput(959.00,412.34)(6.000,2.000){2}{\rule{1.445pt}{0.800pt}}
\put(971,415.34){\rule{2.891pt}{0.800pt}}
\multiput(971.00,414.34)(6.000,2.000){2}{\rule{1.445pt}{0.800pt}}
\put(983,417.84){\rule{2.891pt}{0.800pt}}
\multiput(983.00,416.34)(6.000,3.000){2}{\rule{1.445pt}{0.800pt}}
\put(995,420.84){\rule{2.650pt}{0.800pt}}
\multiput(995.00,419.34)(5.500,3.000){2}{\rule{1.325pt}{0.800pt}}
\put(1006,423.84){\rule{2.891pt}{0.800pt}}
\multiput(1006.00,422.34)(6.000,3.000){2}{\rule{1.445pt}{0.800pt}}
\put(1018,426.84){\rule{2.891pt}{0.800pt}}
\multiput(1018.00,425.34)(6.000,3.000){2}{\rule{1.445pt}{0.800pt}}
\put(1030,429.84){\rule{2.650pt}{0.800pt}}
\multiput(1030.00,428.34)(5.500,3.000){2}{\rule{1.325pt}{0.800pt}}
\put(1041,433.34){\rule{2.600pt}{0.800pt}}
\multiput(1041.00,431.34)(6.604,4.000){2}{\rule{1.300pt}{0.800pt}}
\put(1053,437.34){\rule{2.600pt}{0.800pt}}
\multiput(1053.00,435.34)(6.604,4.000){2}{\rule{1.300pt}{0.800pt}}
\multiput(1065.00,442.38)(1.432,0.560){3}{\rule{1.960pt}{0.135pt}}
\multiput(1065.00,439.34)(6.932,5.000){2}{\rule{0.980pt}{0.800pt}}
\put(1076,446.34){\rule{2.600pt}{0.800pt}}
\multiput(1076.00,444.34)(6.604,4.000){2}{\rule{1.300pt}{0.800pt}}
\multiput(1088.00,451.38)(1.600,0.560){3}{\rule{2.120pt}{0.135pt}}
\multiput(1088.00,448.34)(7.600,5.000){2}{\rule{1.060pt}{0.800pt}}
\multiput(1100.00,456.38)(1.432,0.560){3}{\rule{1.960pt}{0.135pt}}
\multiput(1100.00,453.34)(6.932,5.000){2}{\rule{0.980pt}{0.800pt}}
\multiput(1111.00,461.39)(1.132,0.536){5}{\rule{1.800pt}{0.129pt}}
\multiput(1111.00,458.34)(8.264,6.000){2}{\rule{0.900pt}{0.800pt}}
\multiput(1123.00,467.39)(1.132,0.536){5}{\rule{1.800pt}{0.129pt}}
\multiput(1123.00,464.34)(8.264,6.000){2}{\rule{0.900pt}{0.800pt}}
\multiput(1135.00,473.40)(0.913,0.526){7}{\rule{1.571pt}{0.127pt}}
\multiput(1135.00,470.34)(8.738,7.000){2}{\rule{0.786pt}{0.800pt}}
\multiput(1147.00,480.40)(0.825,0.526){7}{\rule{1.457pt}{0.127pt}}
\multiput(1147.00,477.34)(7.976,7.000){2}{\rule{0.729pt}{0.800pt}}
\multiput(1158.00,487.40)(0.913,0.526){7}{\rule{1.571pt}{0.127pt}}
\multiput(1158.00,484.34)(8.738,7.000){2}{\rule{0.786pt}{0.800pt}}
\multiput(1170.00,494.40)(0.774,0.520){9}{\rule{1.400pt}{0.125pt}}
\multiput(1170.00,491.34)(9.094,8.000){2}{\rule{0.700pt}{0.800pt}}
\multiput(1182.00,502.40)(0.700,0.520){9}{\rule{1.300pt}{0.125pt}}
\multiput(1182.00,499.34)(8.302,8.000){2}{\rule{0.650pt}{0.800pt}}
\multiput(1193.00,510.40)(0.674,0.516){11}{\rule{1.267pt}{0.124pt}}
\multiput(1193.00,507.34)(9.371,9.000){2}{\rule{0.633pt}{0.800pt}}
\multiput(1205.00,519.40)(0.674,0.516){11}{\rule{1.267pt}{0.124pt}}
\multiput(1205.00,516.34)(9.371,9.000){2}{\rule{0.633pt}{0.800pt}}
\multiput(1217.00,528.40)(0.489,0.512){15}{\rule{1.000pt}{0.123pt}}
\multiput(1217.00,525.34)(8.924,11.000){2}{\rule{0.500pt}{0.800pt}}
\multiput(1228.00,539.40)(0.539,0.512){15}{\rule{1.073pt}{0.123pt}}
\multiput(1228.00,536.34)(9.774,11.000){2}{\rule{0.536pt}{0.800pt}}
\multiput(1240.00,550.40)(0.539,0.512){15}{\rule{1.073pt}{0.123pt}}
\multiput(1240.00,547.34)(9.774,11.000){2}{\rule{0.536pt}{0.800pt}}
\multiput(1253.41,560.00)(0.511,0.536){17}{\rule{0.123pt}{1.067pt}}
\multiput(1250.34,560.00)(12.000,10.786){2}{\rule{0.800pt}{0.533pt}}
\multiput(1265.40,573.00)(0.512,0.639){15}{\rule{0.123pt}{1.218pt}}
\multiput(1262.34,573.00)(11.000,11.472){2}{\rule{0.800pt}{0.609pt}}
\multiput(1276.41,587.00)(0.511,0.581){17}{\rule{0.123pt}{1.133pt}}
\multiput(1273.34,587.00)(12.000,11.648){2}{\rule{0.800pt}{0.567pt}}
\multiput(1288.41,601.00)(0.511,0.717){17}{\rule{0.123pt}{1.333pt}}
\multiput(1285.34,601.00)(12.000,14.233){2}{\rule{0.800pt}{0.667pt}}
\multiput(1300.40,618.00)(0.512,0.788){15}{\rule{0.123pt}{1.436pt}}
\multiput(1297.34,618.00)(11.000,14.019){2}{\rule{0.800pt}{0.718pt}}
\multiput(1311.41,635.00)(0.511,0.852){17}{\rule{0.123pt}{1.533pt}}
\multiput(1308.34,635.00)(12.000,16.817){2}{\rule{0.800pt}{0.767pt}}
\multiput(1323.41,655.00)(0.511,0.943){17}{\rule{0.123pt}{1.667pt}}
\multiput(1320.34,655.00)(12.000,18.541){2}{\rule{0.800pt}{0.833pt}}
\multiput(1335.40,677.00)(0.512,1.137){15}{\rule{0.123pt}{1.945pt}}
\multiput(1332.34,677.00)(11.000,19.962){2}{\rule{0.800pt}{0.973pt}}
\multiput(1346.41,701.00)(0.511,1.214){17}{\rule{0.123pt}{2.067pt}}
\multiput(1343.34,701.00)(12.000,23.711){2}{\rule{0.800pt}{1.033pt}}
\multiput(1358.41,729.00)(0.511,1.440){17}{\rule{0.123pt}{2.400pt}}
\multiput(1355.34,729.00)(12.000,28.019){2}{\rule{0.800pt}{1.200pt}}
\multiput(1370.41,762.00)(0.511,1.666){17}{\rule{0.123pt}{2.733pt}}
\multiput(1367.34,762.00)(12.000,32.327){2}{\rule{0.800pt}{1.367pt}}
\multiput(1382.40,800.00)(0.512,2.284){15}{\rule{0.123pt}{3.618pt}}
\multiput(1379.34,800.00)(11.000,39.490){2}{\rule{0.800pt}{1.809pt}}
\put(1391.84,847){\rule{0.800pt}{3.132pt}}
\multiput(1390.34,847.00)(3.000,6.500){2}{\rule{0.800pt}{1.566pt}}
\put(878.0,407.0){\rule[-0.400pt]{2.650pt}{0.800pt}}
\sbox{\plotpoint}{\rule[-0.500pt]{1.000pt}{1.000pt}}%
\put(854,663){\makebox(0,0)[l]{R/L $T=T_L$ }}
\multiput(734,663)(20.756,0.000){5}{\usebox{\plotpoint}}
\put(834,663){\usebox{\plotpoint}}
\put(281,652){\usebox{\plotpoint}}
\put(281.00,652.00){\usebox{\plotpoint}}
\put(301.57,649.44){\usebox{\plotpoint}}
\put(321.04,642.48){\usebox{\plotpoint}}
\put(339.37,632.75){\usebox{\plotpoint}}
\put(356.64,621.24){\usebox{\plotpoint}}
\put(373.49,609.13){\usebox{\plotpoint}}
\put(390.22,596.84){\usebox{\plotpoint}}
\put(406.82,584.38){\usebox{\plotpoint}}
\put(423.05,571.46){\usebox{\plotpoint}}
\put(439.66,559.01){\usebox{\plotpoint}}
\put(456.40,546.74){\usebox{\plotpoint}}
\put(473.67,535.22){\usebox{\plotpoint}}
\put(490.94,523.71){\usebox{\plotpoint}}
\put(508.56,512.76){\usebox{\plotpoint}}
\put(526.90,503.05){\usebox{\plotpoint}}
\put(544.80,492.60){\usebox{\plotpoint}}
\put(563.70,484.07){\usebox{\plotpoint}}
\put(582.39,475.09){\usebox{\plotpoint}}
\put(601.54,467.11){\usebox{\plotpoint}}
\put(620.92,459.69){\usebox{\plotpoint}}
\put(640.61,453.13){\usebox{\plotpoint}}
\put(660.20,446.27){\usebox{\plotpoint}}
\put(680.18,440.68){\usebox{\plotpoint}}
\put(700.26,435.44){\usebox{\plotpoint}}
\put(720.55,431.21){\usebox{\plotpoint}}
\put(740.93,427.35){\usebox{\plotpoint}}
\put(761.52,424.91){\usebox{\plotpoint}}
\put(782.07,422.16){\usebox{\plotpoint}}
\put(802.75,420.39){\usebox{\plotpoint}}
\put(823.43,418.63){\usebox{\plotpoint}}
\put(844.16,418.00){\usebox{\plotpoint}}
\put(864.92,418.00){\usebox{\plotpoint}}
\put(885.67,418.00){\usebox{\plotpoint}}
\put(906.37,419.45){\usebox{\plotpoint}}
\put(927.04,421.25){\usebox{\plotpoint}}
\put(947.73,422.98){\usebox{\plotpoint}}
\put(968.27,425.77){\usebox{\plotpoint}}
\put(988.77,428.96){\usebox{\plotpoint}}
\put(1009.00,433.50){\usebox{\plotpoint}}
\put(1029.28,437.82){\usebox{\plotpoint}}
\put(1049.35,443.09){\usebox{\plotpoint}}
\put(1069.09,449.49){\usebox{\plotpoint}}
\put(1088.71,456.24){\usebox{\plotpoint}}
\put(1108.32,463.03){\usebox{\plotpoint}}
\put(1127.53,470.89){\usebox{\plotpoint}}
\put(1146.69,478.87){\usebox{\plotpoint}}
\put(1165.28,488.03){\usebox{\plotpoint}}
\put(1183.88,497.20){\usebox{\plotpoint}}
\put(1201.90,507.45){\usebox{\plotpoint}}
\put(1219.86,517.82){\usebox{\plotpoint}}
\put(1237.24,529.16){\usebox{\plotpoint}}
\put(1254.51,540.68){\usebox{\plotpoint}}
\put(1271.56,552.50){\usebox{\plotpoint}}
\put(1288.21,564.90){\usebox{\plotpoint}}
\put(1304.62,577.60){\usebox{\plotpoint}}
\put(1321.04,590.28){\usebox{\plotpoint}}
\put(1337.69,602.68){\usebox{\plotpoint}}
\put(1354.37,615.03){\usebox{\plotpoint}}
\put(1371.54,626.69){\usebox{\plotpoint}}
\put(1389.24,637.49){\usebox{\plotpoint}}
\put(1407.98,646.33){\usebox{\plotpoint}}
\put(1428.11,651.09){\usebox{\plotpoint}}
\put(1439,652){\usebox{\plotpoint}}
\end{picture}
\end{center}
\caption{Spatial dependence of the decoherence time. The two top curves 
correspond to a $T=1\ K$ whereas bottom ones correspond
to $T=10\ K$. Here $\theta = 2\pi\sigma_{12}/L$ and
$k_BT_L=2\pi v_S\hbar/L$.}
\label{fig:tau}
\end{figure}

Frequencies of each oscillator get renormalized. As
explained by M\'elin, Dou\c{c}ot and Butaud \cite{Melin:94-1}, such an effect leads
to a kind of ``dynamical orthogonality catastrophe'', that
is to say, a quick decrease of the two fermion Green
function due to phase mixing.

\paragraph{Long time behaviour} The previous
approximation is not valid anymore at longer times
because of dissipation. A Markovian master equation 
{\em \`a la} Caldeira-Legett \cite{CL-83a} can be
obtained for the $l=1$ modes provided the
temperature is bigger than dissipation $k_BT\gg\hbar
\gamma$, where the damping coefficient is given
by $\gamma ={4\pi
\over 3}\,{\alpha_{\mathrm{QED}}\over \alpha}
\,\left({v_S\over c}\right)^2\ldotp {v_S\over L}$. 
In this framework, retaining only
contributions from the $l=1$ modes, we are
able to compute the decoherence coefficient for
$t\rightarrow +\infty$. The short time asymptotics of
this computation coincides with linear terms computed
above. For example the long time dependance of
decoherence at zero temperature is given by:
\begin{eqnarray}
d_{(R/R)}(t,\sigma_1,\sigma_2) & = & 4\Delta_{n,m}\,\ldotp\,
\sin^2{\left({\pi \sigma_{12}\over L}\right)}\ldotp
(1-e^{-\gamma t})\\
d_{(R/L)}(t,\sigma_1,\sigma_2) & = &
2\Delta_{n,m}\,\ldotp\,\left(
1+{m^2-4n^2\alpha^2\over
  m^2+4n^2\alpha^2}
\cos{\left({2\pi\sigma_{12}\over L}\right)}\right)
\ldotp(1-e^{-\gamma t})
\end{eqnarray}
Of course, at non-zero temperature, thermalization
effects enter the game within the dissipation time scale
(damping of hydrodynamic modes) but analytical
computations not reported here can also be performed.

\section{Conclusion}
As a conclusion, let us sketch the decoherence scenario
for Schr\H{o}dinger cats (\ref{defCat}) in a Luttinger
liquid. First of all, at very short times $t\simeq \Lambda^{-1}$, 
the Luttinger parameters $\alpha$ and
$v_S$  get renormalized. Zero modes decohere (only
for $R/L$ and $L/R$ cats) very weakly. Hydrodynamic modes undergo a
gaussian decoherence ($\exp{(-t^2/\tau^2)}$).
After this transitory regime and before saturation, 
the main decoherence contribution 
behaves like $\exp{(-t\gamma(\sigma_1,\sigma_2))}$.
The main contribution for this decoherence time is the first 
hydrodynamic mode.
This decoherence time is of the the order of the
dissipation time, 
{\it i.e.} roughly $10^9$ times longer than the Luttinger
time $L/v_S$.
After the decoherence time scale, decoherence saturates
to its final value. 
However, interference terms are not completely
suppressed. Although our work is done in a many-body
framework, and even in a non-Fermi liquid, it would be
very interesting to make the connection with
Stern {\it et al.} \cite{Stern:90-1} more precise.

\medskip

To conclude, let us stress that 
acoustic 3D phonons are expected to produce a different
scenario. Their 
coupling to electrons is much stronger, the sound velocity 
smaller to $v_S$, and finally the phonon spectrum
presents a natural cutoff at the Debye frequency, which
is much smaller than the Luttinger frequency
$v_S/L$. 
Non-markovian and strong coupling effects should therefore play a much more
important role than in the Luttinger \& QED case. We hope
to come back on these issues in a forthcoming publication.

\section*{Acknowledgments}
We would like to thank the organizers of the conference for giving us
the opportunity to present this work in a very stimulating atmosphere. 
We would also like to 
thank B.~Dou\c{c}ot, Ch.~Chaubet and R.~M\'elin for 
many interesting discussions.
L.~Cugliandolo provided us with useful references on the QBM. During the
conference, we also 
benefited from useful discussions with Th.~Martin and D.~Loss on Luttinger
liquid coupled to QED or phonons.

\medskip

S.~Peysson is supported by the Minist\`ere de l'Education
Nationale, de la Recherche et de la Technologie (AC
contract 98026). This
work also benefited of support from CNRS and the
European Union (TMR FMRX-CT96-0012).

\end{document}